\documentclass[conference]{IEEEtran}
\IEEEoverridecommandlockouts
\usepackage{cite}
\usepackage{amsmath,amssymb,amsfonts}
\usepackage{algorithmic}
\usepackage{graphicx}
\usepackage{textcomp}
\usepackage{xcolor}
\usepackage{balance}

\usepackage[letterpaper,%
            left=0.75in,right=0.75in,top=0.75in,bottom=1in,%
            footskip=.25in,bindingoffset=0.2in]{geometry}

\def\BibTeX{{\rm B\kern-.05em{\sc i\kern-.025em b}\kern-.08em
    T\kern-.1667em\lower.7ex\hbox{E}\kern-.125emX}}

\begin{document}

\makeatletter
\newcommand{\linebreakand}{%
 \end{@IEEEauthorhalign}
 \hfill\mbox{}\par
 \mbox{}\hfill\begin{@IEEEauthorhalign}
}
\makeatother

\title{Short-Term Stock Price Forecasting using exogenous variables and Machine Learning Algorithms

%\thanks{We would like to acknowledge and thank the Post Degree Diploma and the Work on Campus programs at Langara College for supporting our research.}
}

\author{
    \IEEEauthorblockN{Albert Wong}
    \IEEEauthorblockA{\textit{Mathematics and Statistics} \\
    \textit{Langara College}\\
    Vancouver, Canada \\
    0000-0002-0669-4352}
   
   \and 
    \IEEEauthorblockN{Steven Whang}
    \IEEEauthorblockA{\textit{Mathematics and Statistics} \\
    \textit{Langara College}\\
    Vancouver, Canada \\
    swhang00@mylangara.ca}
    
    \and
    \IEEEauthorblockN{Emilio Sagre}
    \IEEEauthorblockA{\textit{Mathematics and Statistics} \\
    \textit{Langara College}\\
    Vancouver, Canada \\
    esagre00@mylangara.ca}

    \linebreakand

   \IEEEauthorblockN{Niha Sachin}
    \IEEEauthorblockA{\textit{Mathematics and Statistics} \\
    \textit{Langara College}\\
    Vancouver, Canada \\
    nsachin00@mylangara.ca}
    
    \and
    \IEEEauthorblockN{Gustavo Dutra}
    \IEEEauthorblockA{\textit{Mathematics and Statistics} \\
    \textit{Langara College}\\
    Vancouver, Canada \\
    gdutra01@mylangara.ca}
    
    \and
    \IEEEauthorblockN{Yew-Wei Lim}
    \IEEEauthorblockA{\textit{Mathematics and Statistics} \\
    \textit{Langara College}\\
    Vancouver, Canada \\
    ywlim@langara.ca}
    
     \linebreakand
    
    \IEEEauthorblockN{Ga\'etan Hains}
    \IEEEauthorblockA{ \textit{LACL} \\
    \textit{Université Paris-Est}\\
    Créteil, France \\
   0000-0002-1687-8091}

\and 
     \IEEEauthorblockN{Youry Khmelevsky}
    \IEEEauthorblockA{\textit{Computer Science} \\
    \textit{Okanagan College}\\
    Kelowna, Canada \\
    0000-0002-6837-3490}

\and 
     \IEEEauthorblockN{Frank Zhang}
    \IEEEauthorblockA{\textit{School of Computing} \\
    \textit{University of the Fraser Valley}\\
   Abbotsford, Canada \\
    0000-0001-7570-9805}
      
}

\maketitle

\begin{abstract}
Creating accurate predictions in the stock market has always been a great challenge in the finance world. With the rise of machine learning as the next level in the forecasting area, this research paper compares four machine learning models and their accuracy in forecasting three well-known stocks traded in the NYSE in the short term over the period from March 2020 to May 2022. We deploy, develop, and tune XGBoost, Random Forest, Multi-layer Perceptron, and Support Vector Regression models and report the models that produce the highest accuracies from our evaluation metrics: RMSE, MAPE, MTT, and MPE. Using a training data set of 240 trading days, we find that XGBoost gives the highest accuracy despite taking longer (up to 10 seconds) to run. Results from this study may improve with the further tuning of the individual parameters or introducing of more exogenous variables.  

\end{abstract}


\section{Introduction}
Machine learning models and algorithms have become increasingly popular over the past few years and will continue to be used even more in the future. Researchers, analysts, and other professionals have used machine learning and incorporated it into our daily lives. From corporations to individuals, machine learning can be applied in a wide range of areas. In this research paper, we explore the use of machine learning models to predict stock market price forecasts. This work extended those completed in \cite{khaidem2016predicting, nabipour2020predicting, wang2022xgboost, vijh2020stock,devadoss2013forecasting} and built on the ideas used in \cite{Wong2021b, Wong2022Estimation}. A more detailed survey of these works is presented in Section II.

Machine learning algorithms use past data to create short-term forecasts of the movement of the chosen stock prices. However, predicting stock market prices is known to be difficult because many factors contribute to the movement of these prices. Well-known theories such as the Efficient Market Hypothesis and Random Walk theory suggest that it is impossible to beat the market consistently. By comparing four different machine learning algorithms, this research aims to determine the model that produces the most accurate prediction of the movement of the chosen stocks. 

There are many possibilities of exogenous variables that can be used for a machine learning algorithm. For the purpose of this research, we have built on our previous study \cite{Wong2022} and include variables that represent short-term interest rate movement (2-year treasury bonds) as well as inflation (the price of gold and the price of crude oil). This is in addition to variables that we believe are central to the prediction of the price of a stock: movements of the overall market (Dow Jones, SI\&P, and NASDAQ indexes) and longer-term interest rate (5 and 10-year treasury bonds).  

Note that our research has focused on price prediction rather than trading efficiency. Our attempt is to see how traditional wisdom about factors that impact the price of a stock, such as historical prices, movement of the broad market, interest rate, inflation, and other exogenous variables, could be incorporated into a machine learning algorithm to produce an accurate forecast. 
We consider that the inclusion of trading strategies, although an ultimate objective, makes it impossible to analyze objectively the quality of price prediction.  In other words, strategies could succeed by ``chance" even with inaccurate price predictions given a favourable enough market situation. 

This work is a continuation of Wong et al. \cite{Wong2022}.

\section{Existing Work}

Traditionally, stock price prediction uses technical and/or fundamental analysis. According to Khaidem et al. \cite{khaidem2016predicting}, using machine learning to forecast stock prices specifically is rather new. Despite the difficulty of the inherent instability of the stock market, there is evidence that applying machine learning techniques elevate the accuracy of price forecast (see, for example, \cite{nabipour2020predicting}). 

As with any other model, the list of predictors (variables) a researcher uses as input into a machine learning algorithm is a critical success factor. Nabipour et al. \cite{nabipour2020predicting} conclude in their research that is utilizing binary data compared to continuous data produces more accurate results when applying machine learning models to the Tehran stock market.

Kim and In \cite{kim2007relationship} suggest a significant relationship between stock prices and bond yield. They conclude that a negative relationship exists between the two variables. Similarly, Engsted and Tanggaard \cite{engsted2001danish} investigate the correlation between the price of stock and bonds in the Danish market. Interestingly, they conclude that the Danish stock and bond prices move similarly to the US market and that the Danish stock and bond prices also have a strong correlation. In his research, Kwan \cite{kwan1996firm} concludes that bond prices and stocks are negatively correlated.

Another variable that may contribute to the movement of stock prices is gold, as it has always been a safe investment in an economic downturn. Smith \cite{smith2001price} explores the movement of the price of gold and stock prices in the United States. The researcher concludes that there is a small but positive correlation between stock prices and gold in the short run. Palamalai and Prakasam \cite{palamalai2015gold} also conclude in their research that there is a long-term relationship between gold and stock prices in the Indian market. 

Similar to gold, another commodity that may have a significant effect on the movement of stocks is crude oil. Narayan \cite{narayan2010modelling} finds that the Vietnamese stock market is heavily dependent on the price of oil. In contrast, international stock markets such as Australia, Canada,  and France do not respond to shocks in the oil market, according to Apergis and Miller \cite{apergis2009structural}. Although this may be true, Akoum et al. \cite{akoum2012co} recently found that the relationship between oil and stock prices has strengthened after interest in oil markets spiked in 2007. 

Machine learning algorithms are increasingly popular as tools for forecasting stock prices. A comprehensive review of this topic can be found in \cite{Joiner2021}. In the following, we highlight some of the basic algorithms used in recent years and are considered in this research.

The Random Forest algorithm is mainly implemented for classification and estimation problems through the use of an ensemble of decision trees. With a goal of minimizing noise originating from the stock market, Vijh et al. \cite{vijh2020stock} apply the random forest algorithm on five well-known stocks traded in the NYSE to forecast their closing prices. They find that artificial neural networks (ANN) perform better than the random forest model. However, Kumar et al. \cite{kumar2018comparative} find that random forest outperforms Support Vector Machine (SVM), K-Nearest Neighbor (KNN), Naive Bayes, and Softmax when predicting stock prices.  

The Support Vector Regression (SVR) algorithm has also been used by researchers to forecast stock prices. Henrique et al. \cite{henrique2018stock} conclude in their research on the Brazil, US, and China markets that the SVR model has predictive power and that the SVR model works better during periods of lower volatility. In this study, the Radial Basis Function (RBF) Kernel is apt because it computes the similarity or how close two points X1 and X2 are to each other due to their similarity to the Gaussian distribution. 

There have been numerous research studies that have used the Multilayer Perceptron (MLP) algorithm to forecast stock prices. Devadoss and Ligori \cite{devadoss2013forecasting} explore the use of MLP on the Bombay Stock Exchange. They conclude that despite the instability and volatility of the market, their MLP model produces accurate results (MAPE ranges from 1.51\% to 5.14\%.) In a more recent study, Namdani and Durrani \cite{namdari2021multilayer} compare the predictive power of MLP to other algorithms such as SVM and Long Short Term Memory (LSTM). They conclude that their MLP forecasts are more accurate than other machine learning models. 

The use of XGBoost to forecast stock prices is a relatively new concept as not many researchers have used it for that purpose. Wang et al. \cite{wang2022xgboost} conclude in their research that XGBoost can accurately predict the prices of stock indices traded in the NYSE. In another study, Gumelar et al. \cite{gumelar2020boosting} explore the use of XGBoost and LSTM to create a trading strategy for the Indonesian Stock Exchange. They conclude that XGBoost produces a 99\% accuracy rate in terms of predictive power. 

\section{Methodology}

The overall process of the research can be seen in Figure 1.  

\begin{figure}[ht]
\centerline{\includegraphics[width=\columnwidth]{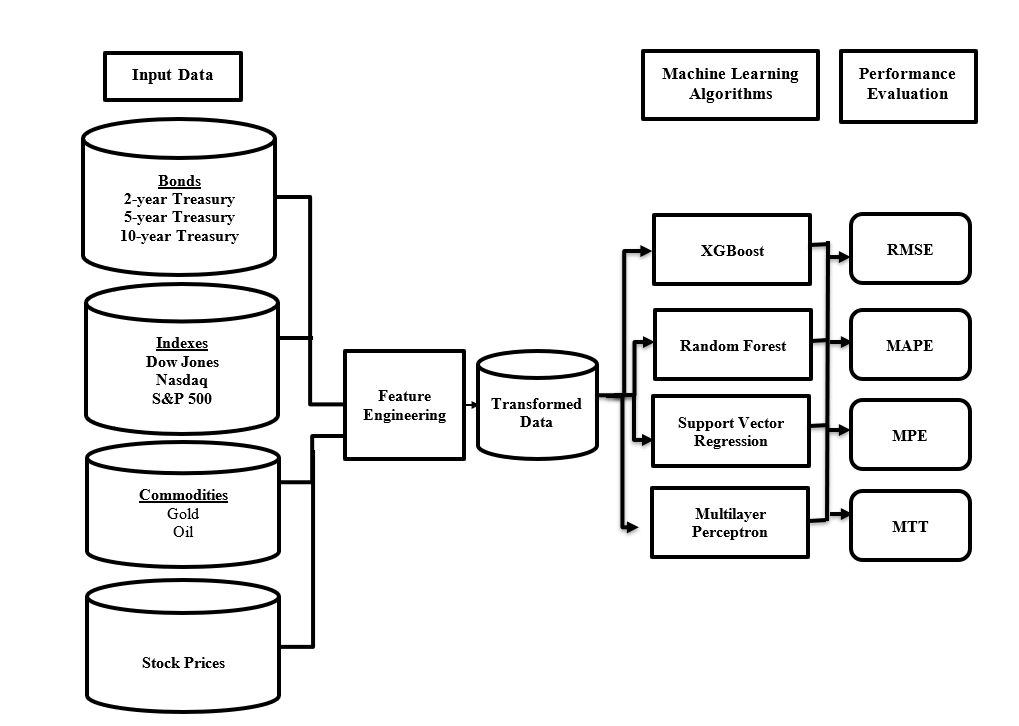}}
\caption{Model Building Process}
\label{fig:OveralProcess}
\end{figure}

Note that the exogenous (input) variables were chosen for their perceived impact on the stock price. Bond prices are related to interest rates, which can affect the borrowing power of investors. Stock market indexes are composed of the ``typical" stocks in the market, so their movement gives the general direction of the market. We use the price of gold and oils as proxies for inflation which we believe have a significant impact on the stock market in general. 

In this study, the forecast models were created using data over 60 days or 240 days and assessed using a testing data set of four months.

To produce forecasts every fifteen minutes during a trading day, a forecasting model should use historical data up until the time period just before the forecasting period. The performance of the model developed would then be evaluated interval by interval over the testing period chosen by the researcher. Therefore, the model will need to be trained and re-trained using the ``rolling” data set (dropping data from one fifteen-minute interval at the ``beginning” interval of the time series and adding data on the most recent  fifteen-minute interval). Forecasts from the model for the next fifteen minutes would then be generated and compared to the actual values. As a result, the training and testing cycle will need to be run with the ``rolling" data set of roughly two thousand four hundred (4 months times 20 trading days each month times 30 trading periods each day i.e. 7.5 hours times 
4 fifteen-minute intervals) times.

We anticipate that the forecasts produced by the models for the next fifteen minutes will be fairly accurate \cite{Wong2023}. On the other hand, the training and test time required to build the model could be very high. Therefore, we made modifications to the above process to reduce the number of training and testing cycles by using a longer forecasting period. Accordingly, we built models using a data set that ``rolls" every day producing one-day ahead forecasts in fifteen-minute intervals as well as one that ``rolls" every five days producing five-day ahead forecasts in fifteen-minute intervals. 

The process described above was used to create a forecasting model for the stock price of Tesla (TSLA). For comparison purposes, we also applied the same approach and algorithms to create similar models for the stock prices of Apple (AAPL), and Nvidia (NVDA).

\subsection{Feature Engineering}

For this study, data have been collected, on the features described in fifteen-minute intervals, for the period of March 2020 to May 2022. In addition to the input numerical features previously mentioned, one-up categorical features were also created to capture the calendar effects and seasonality patterns on stock prices.  A summary of the variables is presented below.

\subsubsection{Numerical Features}
\begin{itemize}
 \item Value of Dow Jones Index
 \item Value of Nasdaq Index
 \item Value of S\&P 500 Index
 \item Price of two-year treasury bond
 \item Price of Five-year treasury bond
 \item Price of Ten-year treasury bond
 \item Price of gold
 \item Price of crude oil
 
\end{itemize}

The value of these numerical features was shifted by one time period so that the stock prices were predicted based on the value of these features in the previous period.

\subsubsection{Categorical (One-Up) Features}

To capture the seasonal pattern and other calendar effects on stock prices, we created several indicator features for each fifteen-minute interval: 

\begin{itemize}
    \item Months of the year (12 one-hot variables)
    \item Day of the month (31 one-hot variables)
    \item Day of the Week (5 one-hot variables for Monday to Friday)
    \item Hours of the day (6 one-hot variables for hours 9:00 to 16:00)
    \item Minute Segment of the hour ( 4 one-hot variables for minute segment between 0,15,30, and 45)
    \item Whether the time period is on Monday morning (1 indicator variable)
    \item Whether the time period is on Friday afternoon (1 indicator variable)
    \item Whether the time period is in a ``Pre-holiday" afternoon (1 indicator variable)
    \item Whether the time period is in a ``post-holiday" morning (1 indicator variable)
\end{itemize}

\subsection{Normalization of Numerical Features}

The min-max normalization process (Equation 1) is applied across all numerical features.  

\begin{equation}
    X_{normalized} = \frac{X-X_{min}}{X_{max}-X_{min}}
    \label{eq:minmax}
\end{equation}

\subsection{Performance Evaluation}

After building the machine learning models, we must evaluate how accurate the forecasts are. In this research, we use the following four performance metrics: Root mean squared error, Mean absolute percentage error, Mean positive error, and Mean training time. The first two performance metrics are common in estimation or forecasting models and would allow for the comparison of results across models or studies. \cite{Wong2022Estimation,Wong2022}

\subsubsection{Root Mean Square Error}
The root mean square error (RMSE) is a popular metric for measuring the predictive model's performance and comparing different predictive models. The calculation for finding the value of RMSE can be seen in Equation 2 below. 

\begin{equation}
RMSE = \sqrt{\frac{1}{n} \sum_{i=1}^{n}(Y_i-\hat{Y_i})^2}
\end{equation}
In this project, the unit of the RMSE is in dollars. 

\subsubsection{Mean Absolute Percentage Error}
The mean absolute percentage error (MAPE) metric measures the average absolute error percentage between the predicted and the actual value. It is independent of the scale of the data and therefore can be used for  comparison of performance between different models. The calculation to find MAPE can be seen below in Equation 3. 

\begin{equation}
MAPE = \frac{1}{n} {\sum_{i=1}^{n} \frac{|Y_i-\hat{Y_i}|}{|Y_i|}}
\label{eq:mape}
\end{equation}

\subsubsection{Mean Positive Error}

The mean positive error (MPE) is developed specifically for the evaluation of stock price prediction. It focuses solely on errors made when the predicted value is larger than the actual value.  The calculation to find MPE can be seen below in equation 4.\ref{eq:mpe}.

\begin{equation}
MPE = \frac{1}{n} {\sum_{i=1}^{n} max(\hat{Y_i}-Y_i,0)}
\label{eq:mpe}
\end{equation}

The unit of the MPE is in dollars. 

\subsubsection{Mean Training Time}

The Mean Training Time (MTT) measures the average amount of time an algorithm takes to train with the dataset and create the prediction model. From a computational standpoint, the development process described above could be demanding if we were to produce forecasts every fifteen minutes for the entire stock universe in a sizable stock market. Therefore, the mean training time is tracked as part of the evaluation metric. 

 As you can see in the following section, this metric is not critical for the current study as the training time for the development of a forecasting model for one stock is insignificant. This measure will become critical when training a model that allows for the forecasting of multiple stocks.

The unit of the MTT is in seconds. 

\section{Results}

In this section, we compare the performance of the models developed in this study based on the evaluation metrics mentioned above. Naturally, a lower value for all performance metrics is favourable for a model as this implies that its predictions are closer to the actual values in general. In the following, we will focus our comparisons on the use of a 60 days training data set versus a 240 days training data set for producing one-day ahead and five-day ahead forecasts. 

The four basic machine learning algorithms were used to develop the forecasting models with different hyper-parameters to optimize their performance. The following tables summarize the top-performing models developed for each algorithm with details on the hyper-parameters used. 

\begin{figure}[ht]
\centerline{\includegraphics[width=\columnwidth]{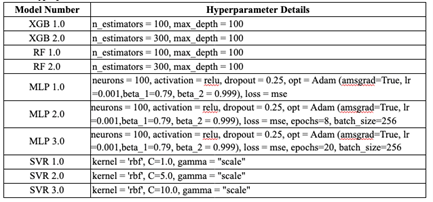}}
\caption{Top-Performing Models for Each ML Algorithm}
\label{fig: Model Hyperparameters}
\end{figure}

The first set of results, presented in Figure 3, is on the comparison of these models for forecasting the price of Tesla stock for the next day (1 day ahead forecast). Two sets of results were produced: one for models developed and evaluated using ``rolling” 60-day training data sets, the other using ``rolling” 240-day training data sets.

It is apparent that, on average, the XGBoost models produce the lowest errors compared to the other machine learning models considered. The Random Forest models are not far behind.

\begin{figure}[ht]
\centerline{\includegraphics[width=\columnwidth]{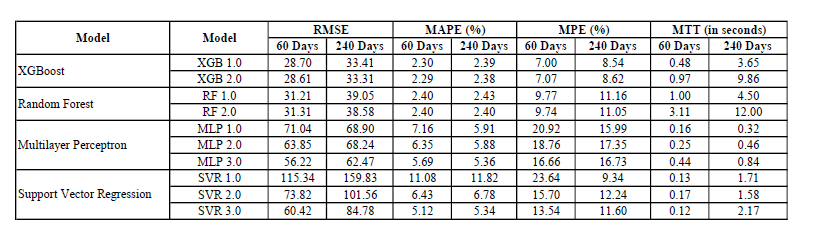}}
\caption{Performance of Forecasting Models on Price of Tesla Stock - One Day Ahead}
\label{fig:  Telsa - 1 day ahead Forecast}
\end{figure}

The same can be said about the performance of forecasting models developed for the Apple and Nvidia stocks.

\begin{figure}[ht]
\centerline{\includegraphics[width=\columnwidth]{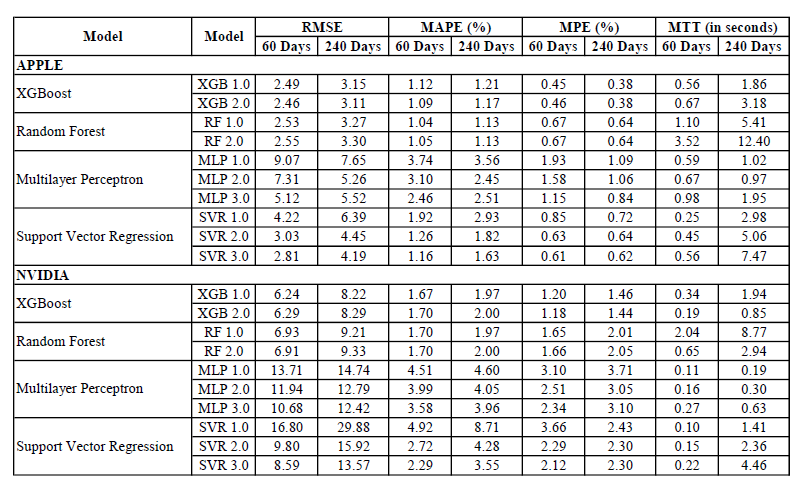}}
\caption{Performance of Forecasting Models on Price of Apple and Nvidia - One Day Ahead}
\label{fig:  Other - 1 day ahead Forecast}
\end{figure}

To give the reader a sense of the performance of the models, below are the graphs that show the best models, in terms of MAPE, in producing 1 day-ahead forecast for the three stocks.

\begin{figure}[ht]
\centerline{\includegraphics[width=\columnwidth]{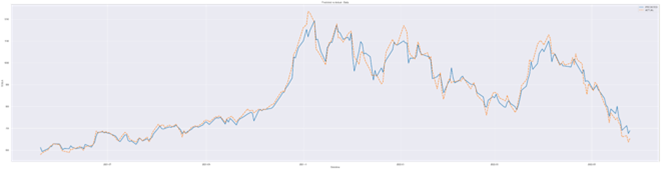}}
\caption{Forecasting 1-Day ahead using 240 training days for Tesla using XGBoost}
\label{fig: XGBOOST Predicting Tesla stock (240 training days) - 1 Day ahead }
\end{figure}

\begin{figure}[ht]
\centerline{\includegraphics[width=\columnwidth]{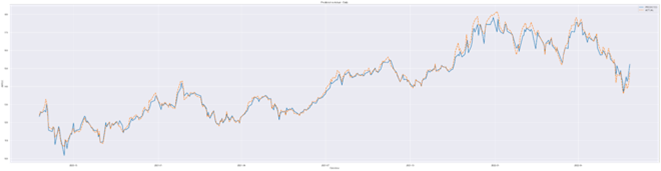}}
\caption{Forecasting 1-Day ahead using 240 training days for Apple using XGBoost}
\label{fig: XGBOOST Predicting Apple stock (240 training days) - 1 Day ahead}
\end{figure}

\begin{figure}[ht]
\centerline{\includegraphics[width=\columnwidth]{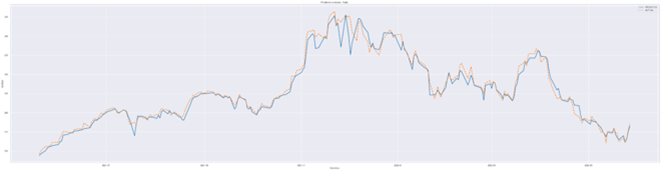}}
\caption{Forecasting 1-Day ahead using 240 training days for Nvidia using XGBoost }
\label{fig: XGBOOST Predicting Nvidia stock (240 training days) - 1 Day ahead }
\end{figure}

It can be observed from the graphs that prediction accuracy is higher during periods with low volatility. A higher level of prediction errors occurs when the actual price of the stocks fluctuates more. This is completely understandable.

Across all experiments, the XGBoost models produced the lowest errors on average compared to the other machine learning models. However, it is noticeable from the MTT metric that the Random Forest algorithm takes the longest to train, followed by XGBoost. More interesting is the fact that increasing the N-estimators parameter from 100 to 300 for both the XGBoost and the Random Forest algorithm showed little to no signs of improvement with respect to the MAPE performance measure. This is true regardless of whether the training data set is for 60 days or 240 days. 

The second set of results, presented in Figure 8, is on the comparison of these models for forecasting the price of all three stocks for the next five days (5-day ahead forecast). As before, two sets of results were produced: one for models developed and evaluated using ``rolling” 60-day training data sets, the other using ``rolling” 240-day training data sets.

\begin{figure}[ht]
\centerline{\includegraphics[width=\columnwidth]{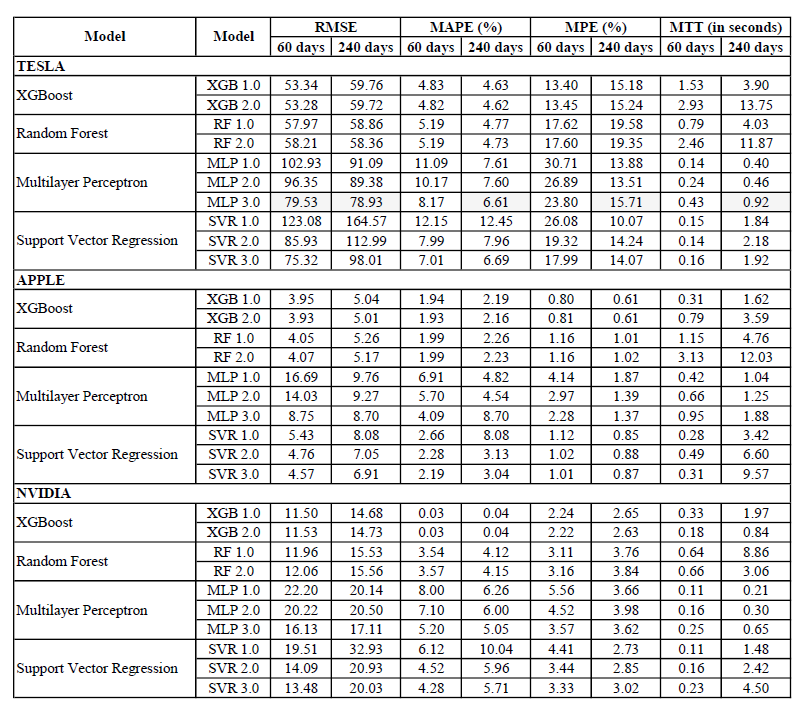}}
\caption{Performance of Forecasting Models - Five Days Ahead}
\label{fig: 5 days ahead Forecast}
\end{figure}

Overall, all values of performance metrics considered for the 5-day ahead forecasts are higher than those for the 1-day ahead forecasts. This outcome is expected as it is more difficult to accurately forecast further into the future, thus errors on average are higher. However, observations on the 5-day ahead forecasts are similar to those presented for the 1-day ahead forecasts. 

With MAPE at 5\% or lower, these forecasting models demonstrated the usefulness of the approach taken in this research. Further work will be required to make these models implementable inside a profitable algorithmic trading system.

\section{Future Work}

Although the results have been insightful, a number of improvements could be made to these models and worked on in the next phase of the research:

\begin{itemize}
    \item the development of a data depository or data house to acquire, store, and manage the volume of stock price and related data over time. Data in this facility will be the data source for the research proposed below.
    \item expansion of the stock universe being considered in the research.
    \item additional features on the economic environment. Examples would be the inflation rate, unemployment rate, etc.
     \item features on market sentiment through data from social media and financial-related forums, etc.
    \item demographic features related to the companies: industrial classification, size, profitability, etc.
    \item further tuning of the above models on the hyper-parameters.
    \item consideration of other training data periods
    \item monitoring of the mean training time as the models become more complicated as a result of the above.
\end{itemize}

\section{Conclusions}

This study explores the predictive power of four different machine learning algorithms (XGBoost, Random Forest, Support Vector Regression, and Multilayer Perceptron) and how accurately they can forecast the prices of several technology stocks (Apple, Nvidia, and Tesla). As well, this research follows the ``fundamental" approach in finance for evaluating the stock market and picked features as input to the models accordingly. The results of these models showed promise as the values on several performance metrics are quite reasonable. We believe that the extension of these models to include other economic- or financial-related measures such as inflation rate and company demographics will lead to a stock forecasting model as the foundation of a profitable algorithmic trading system.

\bibliographystyle{IEEEtran}
\balance
\bibliography{references.bib}

\begin{thebibliography}{10}
\providecommand{\url}[1]{#1}
\csname url@samestyle\endcsname
\providecommand{\newblock}{\relax}
\providecommand{\bibinfo}[2]{#2}
\providecommand{\BIBentrySTDinterwordspacing}{\spaceskip=0pt\relax}
\providecommand{\BIBentryALTinterwordstretchfactor}{4}
\providecommand{\BIBentryALTinterwordspacing}{\spaceskip=\fontdimen2\font plus
\BIBentryALTinterwordstretchfactor\fontdimen3\font minus
  \fontdimen4\font\relax}
\providecommand{\BIBforeignlanguage}[2]{{%
\expandafter\ifx\csname l@#1\endcsname\relax
\typeout{** WARNING: IEEEtran.bst: No hyphenation pattern has been}%
\typeout{** loaded for the language `#1'. Using the pattern for}%
\typeout{** the default language instead.}%
\else
\language=\csname l@#1\endcsname
\fi
#2}}
\providecommand{\BIBdecl}{\relax}
\BIBdecl

\bibitem{khaidem2016predicting}
L.~Khaidem, S.~Saha, and S.~R. Dey, ``Predicting the direction of stock market
  prices using random forest,'' \emph{arXiv preprint arXiv:1605.00003}, 2016.

\bibitem{nabipour2020predicting}
M.~Nabipour, P.~Nayyeri, H.~Jabani, S.~Shahab, and A.~Mosavi, ``Predicting
  stock market trends using machine learning and deep learning algorithms via
  continuous and binary data; a comparative analysis,'' \emph{IEEE Access},
  vol.~8, pp. 150\,199--150\,212, 2020.

\bibitem{wang2022xgboost}
J.~Wang, Q.~Cheng, and Y.~Dong, ``An xgboost-based multivariate deep learning
  framework for stock index futures price forecasting,'' \emph{Kybernetes}, no.
  ahead-of-print, 2022.

\bibitem{vijh2020stock}
M.~Vijh, D.~Chandola, V.~A. Tikkiwal, and A.~Kumar, ``Stock closing price
  prediction using machine learning techniques,'' \emph{Procedia computer
  science}, vol. 167, pp. 599--606, 2020.

\bibitem{devadoss2013forecasting}
A.~V. Devadoss and T.~A.~A. Ligori, ``Forecasting of stock prices using multi
  layer perceptron,'' \emph{International journal of computing algorithm},
  vol.~2, no.~1, pp. 440--449, 2013.

\bibitem{Wong2021b}
A.~Wong, D.~Joiner, C.~Chiu, M.~Elsayed, K.~Pereira, Y.~Khmelevsky, and
  J.~Mahony, ``{A Survey of Natural Language Processing Implementation for Data
  Query Systems},'' in \emph{2021 IEEE International Conference on Recent
  Advances in Systems Science and Engineering (RASSE)}, 2021, pp. 1--8.

\bibitem{Wong2022Estimation}
{A. Wong}, {C. Chiu}, {A. Abdulgapul}, {M. N. Beg}, {Y. Khmelevsky}, and {J.
  Mahony}, ``{Estimation of Hourly Utility Usage Using Machine Learning},'' in
  \emph{IEEE International Systems Conference (SysCon) 2022}, 2022.

\bibitem{Wong2022}
A.~Wong, J.~Figini, A.~Raheem, G.~Hains, Y.~Khmelevsky, and P.~C. Chu,
  ``Forecasting of stock prices using machine learning models,''
  \emph{Submitted}, 2022.

\bibitem{kim2007relationship}
S.~Kim and F.~In, ``On the relationship between changes in stock prices and
  bond yields in the g7 countries: Wavelet analysis,'' \emph{Journal of
  International Financial Markets, Institutions and Money}, vol.~17, no.~2, pp.
  167--179, 2007.

\bibitem{engsted2001danish}
T.~Engsted and C.~Tanggaard, ``The danish stock and bond markets: Comovement,
  return predictability and variance decomposition,'' \emph{Journal of
  Empirical Finance}, vol.~8, no.~3, pp. 243--271, 2001.

\bibitem{kwan1996firm}
S.~H. Kwan, ``Firm-specific information and the correlation between individual
  stocks and bonds,'' \emph{Journal of financial economics}, vol.~40, no.~1,
  pp. 63--80, 1996.

\bibitem{smith2001price}
G.~Smith, ``The price of gold and stock price indices for the united states,''
  \emph{The World Gold Council}, vol.~8, no.~1, pp. 1--16, 2001.

\bibitem{palamalai2015gold}
S.~Palamalai and K.~Prakasam, ``Gold price, stock price and exchange rate
  nexus: The case of india,'' \emph{Srinivasan P. and Karthigai, P.(2014), Gold
  Price, Stock Price and Exchange Rate Nexus: The Case of India, The IUP
  Journal of Financial Risk Management}, vol.~11, no.~3, pp. 1--12, 2015.

\bibitem{narayan2010modelling}
P.~K. Narayan and S.~Narayan, ``Modelling the impact of oil prices on
  vietnam’s stock prices,'' \emph{Applied energy}, vol.~87, no.~1, pp.
  356--361, 2010.

\bibitem{apergis2009structural}
N.~Apergis and S.~M. Miller, ``Do structural oil-market shocks affect stock
  prices?'' \emph{Energy economics}, vol.~31, no.~4, pp. 569--575, 2009.

\bibitem{akoum2012co}
I.~Akoum, M.~Graham, J.~Kivihaho, J.~Nikkinen, and M.~Omran, ``Co-movement of
  oil and stock prices in the gcc region: A wavelet analysis,'' \emph{The
  Quarterly Review of Economics and Finance}, vol.~52, no.~4, pp. 385--394,
  2012.

\bibitem{Joiner2021}
D.~Joiner, A.~Vezeau, G.~Hains, Y.~Khmelevsky, and A.~Wong, ``Algorithmic
  trading and short-term forecast for financial time series with machine
  learning models; state of the art and perspectives,'' \emph{2022 IEEE
  International Conferences on Recent Advances in Systems Science and
  Engineering}, 2022.

\bibitem{kumar2018comparative}
I.~Kumar, K.~Dogra, C.~Utreja, and P.~Yadav, ``A comparative study of
  supervised machine learning algorithms for stock market trend prediction,''
  in \emph{2018 Second International Conference on Inventive Communication and
  Computational Technologies (ICICCT)}.\hskip 1em plus 0.5em minus 0.4em\relax
  IEEE, 2018, pp. 1003--1007.

\bibitem{henrique2018stock}
B.~M. Henrique, V.~A. Sobreiro, and H.~Kimura, ``Stock price prediction using
  support vector regression on daily and up to the minute prices,'' \emph{The
  Journal of finance and data science}, vol.~4, no.~3, pp. 183--201, 2018.

\bibitem{namdari2021multilayer}
A.~Namdari and T.~S. Durrani, ``A multilayer feedforward perceptron model in
  neural networks for predicting stock market short-term trends,'' in
  \emph{Operations Research Forum}, vol.~2, no.~3.\hskip 1em plus 0.5em minus
  0.4em\relax Springer, 2021, pp. 1--30.

\bibitem{gumelar2020boosting}
A.~B. Gumelar, H.~Setyorini, D.~P. Adi, S.~Nilowardono, A.~Widodo, A.~T.
  Wibowo, M.~T. Sulistyono, E.~Christine \emph{et~al.}, ``Boosting the accuracy
  of stock market prediction using xgboost and long short-term memory,'' in
  \emph{2020 International Seminar on Application for Technology of Information
  and Communication (iSemantic)}.\hskip 1em plus 0.5em minus 0.4em\relax IEEE,
  2020, pp. 609--613.

\bibitem{Wong2023}
{A. Wong}, {J. Figini}, {A. Raheem}, {G. Hains}, {Y. Khmelevsky}, and {P. Chu},
  ``{Forecasting of Stock Prices Using Machine Learning Models},'' in
  \emph{IEEE International Systems Conference (SysCon) 2023}, 2023.

\end{thebibliography}

\end{document}